# ADOPTION AND EFFECTIVENESS OF AI-BASED ANOMALY DETECTION FOR CROSS-PROVIDER HEALTH DATA EXCHANGE

**Cao Tram Anh Hoang**

School of Information Systems, Queensland University of Technology, Australia
Email: caotramanh.hoang@hdr.qut.edu.au

# Adoption and Effectiveness of AI-Based Anomaly Detection for Cross-Provider Health Data Exchange

**Abstract**

Cross-provider exchange of electronic health records (EHRs) remains fragmented, creating blind spots for inappropriate access, insider misuse, and privacy breaches. Research on anomaly detection has largely focused on single-site datasets and model accuracy, while overlooking organisational readiness and the contextual audit fields required to make detection actionable across providers.

This study pursues two aims: (1) to identify the digital capabilities healthcare organisations need to adopt AI-based anomaly detection in a cross-provider setting; and (2) to evaluate the effectiveness and operational trade-offs of lightweight, interpretable detection approaches when contextual audit features are available.

A semi-systematic scoping synthesis distils adoption requirements into four domains—Governance, Infrastructure/Interoperability, Workforce, and AI Integration—and formalises them as a 10-item readiness checklist with maturity bands and concrete evidence artefacts. Complementing this, a rapid evidence review is paired with a small cross-provider audit-log simulation. The simulation incorporates contextual fields (provider mismatch, hour-of-day, days-since-discharge, session duration, recent access). Transparent rules are benchmarked against Isolation Forest, and SHAP is used to explain model drivers and interactions.

The readiness checklist operationalises adoption by specifying standard audit fields, role-based access controls, escalation and retraining triggers, and governance thresholds. In the refined dataset (500 sessions; 99 injected anomalies), rules achieved recall of 1.00 with modest false positives (precision ≈0.85), whereas Isolation Forest reduced alert volume but at lower recall. SHAP analyses highlight provider-mismatch combined with off-hours access as the dominant interaction, alongside effects from recent access frequency and role–action patterns.

The paper proposes a staged rollout: standardise audit fields and thresholds; pilot rules to guarantee coverage; layer Isolation Forest to prioritise reviews; embed SHAP-backed explanations and periodic recalibration. Contributions include a readiness instrument, feature-level evidence to guide field standards, and an actionable deployment sequence for health information exchanges and multi-site providers.

## 1. Introduction

The exchange of electronic health records (EHRs) across care providers aims to support continuity of care, reduce duplicate tests, and improve patient safety. Digital health, encompassing information and communication technologies to enhance human health and healthcare delivery (Agarwal et al., 2010), has accelerated digital integration, requiring professionals to adapt to new technologies (Stoumpos et al., 2023). However, the ecosystem remains fragmented, with data flowing through separate systems governed by different organisations, limited interoperability, inconsistent auditing, and weak oversight. Clinicians report mixed EHR experiences, viewing them as efficient yet time-consuming, with concerns over data accuracy and integration (Upadhyay & Hu, 2022). These gaps create blind spots for inappropriate access, insider misuse, and privacy breaches, especially in multi-provider care.

### 1.1. Background: anomaly detection and audit logs

To mitigate misuse of EHR data, researchers have increasingly turned to anomaly detection. Early work relied on rule-based monitoring and collaborative filtering of audit logs to identify unusual access patterns. Menon et al. (2014) used collaborative filtering to detect inappropriate EHR access by comparing users' actions against peer behaviour. Fabbri and LeFevre (2011) proposed "explanation-based auditing," which automatically generates explanations for why an EHR was accessed, aiming to help patients and auditors understand the context. More recent studies deploy machine learning, including Isolation Forest and Local Outlier Factor, to detect anomalies in audit logs (Tabassum et al., 2024). Tabassum et al. showed that Isolation Forest achieved high precision in smart-health systems but acknowledged that most evaluations are limited to single providers and focus on model accuracy rather than organisational deployability. Deep-learning approaches such as EHR-BERT treat sequences of clinical events as tokens to identify abnormal access patterns; although promising, these models can be computationally expensive and often lack interpretability (Niu et al., 2024).

Audit logs—metadata that record who accessed a chart, what action they took, and when—are foundational to anomaly-detection approaches. The HIPAA Security Rule requires such logging (Rule et al., 2020); however, that review also shows most audit-log studies lack standardised measures and contextual detail. Contextual fields—clinician role, any mismatch between requesting provider and record owner, time of access, and days since discharge—could materially improve detection yet are seldom used. Moreover, the literature rarely considers cross-provider data flows or the organisational capabilities needed to operate AI-based monitoring across shared environments.

### 1.2. Research gap

Across reviews, three issues recur. Most anomaly-detection evaluations remain single-provider, offering little insight into performance in multi-organisation settings. Readiness for AI-enabled monitoring is under-theorised: EHR implementation evidence highlights governance, leadership and workforce development (Upadhyay & Hu, 2022), and AI-specific reviews call for structured, healthcare-tailored frameworks spanning leadership, infrastructure, data quality and change management (Nair et al., 2025). Complementing this, digital-readiness evidence shows that usability, training and institutional support are necessary precursors to effective adoption (Alotaibi et al., 2025).

A recent scoping review further underscores trust, regulatory clarity, and organisational culture as central enablers of AI uptake in healthcare (Hassan et al., 2024), yet this literature stops short of linking readiness factors to anomaly-detection systems or to cross-provider data exchange. A third gap concerns the balance between performance, interpretability, and operational feasibility: high-capacity models (e.g., EHR-BERT) may be impractical for routine audits, whereas simple rules are transparent but can inflate false positives. Emerging work calls for approaches that explicitly manage this interpretability–accuracy trade-off and for mechanisms—such as SHAP-based explanations—that render decisions reviewable in practice (Niu et al., 2024).

### 1.3. Objectives and scope

To address these gaps, this paper pursues two objectives:

**RQ1:** What organisational and digital capabilities are required for healthcare providers to effectively adopt AI-based anomaly detection tools across shared data environments and service providers?

**RQ2:** To what extent can AI-based anomaly detection systems accurately and explainably identify suspicious activity in real-time cross-provider health data exchanges?

Cross-provider settings involve patient EHR access by multiple organisations (e.g., hospitals, clinics) in shared exchanges. The scope focuses on audit logs and key contextual features (e.g., user/provider IDs, session details), targeting inappropriate access detection with deployable AI balancing interpretability and performance.

### 1.4. Research approach

RQ1 employs a semi-systematic scoping review with framework synthesis, screening 2011–2025 literature on EHRs, anomaly detection, and AI adoption via PRISMA. Themes in governance, infrastructure/interoperability, workforce, and AI integration form a four-pillar readiness checklist, aligned with frameworks like UTAUT.

RQ2 uses an evidence-informed simulation: Synthetic cross-provider audit logs with injected anomalies evaluate interpretable rules against Isolation Forest (Tabassum et al., 2024), using metrics like precision/recall and SHAP explanations. This benchmarks models while validating checklist prerequisites.

### 1.5. Expected contributions

This study presents three contributions. First, the readiness checklist operationalises adoption requirements for AI-based anomaly detection in cross-provider settings, translating concepts into measurable controls and maturity levels as a practical tool. Second, the simulation offers empirical evidence on effectiveness and interpretability trade-offs between rules and Isolation Forest, highlighting contextual influences like provider mismatch. Third, linking findings to the checklist provides a staged pathway—standardise fields, pilot rules under governance, layer Isolation Forest, and embed explainability—that bridges research and practice.

### 1.6. Paper structure

The remainder of this paper is organised as follows. The Literature Review synthesises research on anomaly detection, audit-log analytics, and AI adoption frameworks, highlighting gaps that motivate this study. The Methodology section details the scoping review and simulation designs, including search strategies, data sources, feature engineering, and evaluation metrics. The Results and Discussion present the readiness checklist, simulation performance, and SHAP-based insights, contextualising them against prior work and discussing implications for practitioners and policymakers. Finally, the Conclusion summarises key contributions, acknowledges limitations, and suggests future research directions.

## 2. Literature Review

This literature review surveys research relevant to AI-enabled anomaly detection in EHRs and the associated organisational factors that influence adoption in cross-provider settings. The review covers peer-reviewed studies published between 2011 and 2025 and is organised by themes rather than strictly chronological order. The first theme summarises seminal approaches for auditing and anomaly detection in EHR access logs. The second theme reviews contemporary machine-learning and deep-learning techniques for EHR anomaly detection. The third theme explores how data fragmentation and contextual metadata affect anomaly detection across multiple providers, including recent efforts to create interoperable knowledge infrastructure. The fourth theme synthesises research on organisational readiness, digital capability and AI adoption frameworks which underpin the deployment of anomaly detection systems. The conclusion highlights gaps in current knowledge and positions the present study.

### 2.1. Early auditing and anomaly-detection approaches

Early work on identifying inappropriate EHR access drew from auditing theory and recommendation systems. Fabbri and LeFevre's (2011) explanation-based auditing framework generated natural language explanations for access logs, demonstrating that such explanations could account for over 94% of accesses. This work emphasised that simple queries rarely provide sufficient context for patients or auditors to understand record access; explanation templates derived from workflows can thus enhance transparency and trust. Around the same time, Menon et al. (2014) employed collaborative filtering to detect inappropriate access by modeling user-patient relationships and comparing observed behaviors with inferred patterns from similar users. This approach showed promise for flagging outliers but relied on historical single-site data and omitted contextual features such as provider affiliation or time of day. These auditing methods aligned with regulatory requirements, including the US HIPAA mandate that hospitals maintain audit logs capturing who accessed a patient's record, what actions occurred, and when. A systematic review by Rule et al. (2020) found that audit logs are increasingly used to study clinical activity, yet many studies lack standardised measures and contextual information for interpreting access events. Early research thus highlighted the potential of audit logs for misuse detection while revealing key gaps: limited context in logs, a focus on single-provider environments, and the absence of validated metrics.

### 2.2. Contemporary machine-learning and deep-learning approaches

Recent years have seen a surge in machine-learning (ML) methods for detecting anomalous access patterns. Tabassum et al. (2024) applied unsupervised algorithms, including Isolation Forest and Local Outlier Factor, to synthetic health-care data and found that these methods achieved high detection rates for insider misuse. Their study emphasised that anomalies often arise from credential theft, negligence or malicious insiders and that ML techniques can help alert security teams. Similarly, Niu et al. (2024) proposed EHR-BERT, a transformer-based model that uses sequences of clinical events and claim codes to detect anomalies and showed that it reduces false positives compared with traditional methods. However, deep models like EHR-BERT require significant computational resources and large labelled datasets; they also operate as “black boxes,” making it difficult for compliance teams to explain why an alert was generated.

More recently, researchers have experimented with multi-modal and diffusion-based models that integrate imaging data. For example, Kim et al. (2024) combined chest-X-ray images with EHR variables using a diffusion model to detect anomalies; yet such approaches remain preliminary and often focus on clinical imaging rather than audit logs. Tang et al. explored multimodal foundation models for medical anomaly detection but did not address the organisational processes needed to implement them. Overall, contemporary ML approaches advance the state of the art in accuracy but still rely on single-site datasets and omit organisational readiness considerations.

### 2.3. Data fragmentation, contextual metadata and cross-provider challenges

EHR interoperability remains a persistent challenge. Clinicians have expressed mixed views about EHRs: while some regard them as beneficial for evidence-based care, others find them time-consuming and worry about data accuracy and incomplete information (Upadhyay & Hu, 2022). Specifically, nurses appreciated EHRs for improving efficiency, whereas other clinicians viewed them as burdensome and were concerned about data accuracy and limited system integration. Inadequate integration leads to fragmented records, forcing clinicians to order duplicate tests or make decisions without complete history.

Several studies highlight how fragmentation undermines anomaly detection and clinical decision support. Shang et al.'s (2024) knowledge-graph study noted that patients often visit multiple hospitals, resulting in fragmented data across institutions. The authors designed an EHR-oriented knowledge graph using the Observational Medical Outcomes Partnership (OMOP) vocabulary and blockchain to share intermediate reasoning results across hospitals while preserving data privacy. Evaluated on chronic kidney disease patients, the system identified 124 patients with CKD by combining fragmented data, whereas single-hospital data alone failed to detect the condition. These findings illustrate that cross-provider collaboration can significantly improve detection of health conditions and anomalies but requires standardised semantics and privacy-preserving data-sharing mechanisms.

Beyond knowledge graphs, other technologies aim to address interoperability. Ferreira et al. (2024) reviewed the application of distributed ledger technology (DLT) to EHR interoperability and security, arguing that current EHR exchange is fragmented and prone to security issues. DLT can create immutable logs, enhance transparency, and allow patients to control access. However, adoption requires standardised protocols and integration with existing systems. Such infrastructure also demands organisational readiness and governance.

### 2.4. Organisational readiness and AI adoption frameworks

The success of AI initiatives in healthcare depends not only on technical performance but also on organisational preparedness. Several studies draw on the UTAUT and institutional readiness frameworks. Alotaibi et al.'s (2025) systematic review found that digital readiness hinges on performance expectancy (perceived benefit), effort expectancy (ease of use), facilitating conditions (training, infrastructure, and support), and social influence; complexity and multiple logins are notable barriers. Do Nascimento et al.'s (2023) umbrella review of digital health adoption across 108 systematic reviews revealed that common barriers include infrastructure and technical limitations, psychological resistance, and increased workload, whereas facilitators include training, multisector incentives, and perceived technology effectiveness.

Nair et al. (2025) observed that healthcare organisations lack structured frameworks for AI implementation and proposed a process framework encompassing planning, execution, and continuous evaluation, emphasising stakeholder engagement and governance as critical for adoption. Hassan et al.'s (2024) scoping review of AI adoption identified barriers such as lack of trust, unclear regulatory frameworks, and poor governance, arguing that building trust through transparency and proper governance is essential. Gardner et al.'s (2023) qualitative study of a UK National Health Service hospital found that institutional readiness depends on organisational vision, technical capability, agility, and talent; unrealistic expectations and resource constraints hinder digital transformation. Together, these studies underline that governance, infrastructure, workforce capability, and integration of AI into existing workflows are prerequisites for successful implementation.

### 2.5. Synthesis and identified gaps

This review shows that anomaly-detection research has evolved from rule-based and collaborative filtering approaches to sophisticated ML and deep-learning methods. While modern models deliver high accuracy, most studies evaluate them in single-provider environments and do not consider the organisational and interoperability context. Meanwhile, research on digital readiness and AI adoption emphasises the need for governance, infrastructure, workforce capability and continuous evaluation, but these works rarely link readiness to the technical design of anomaly-detection systems. The literature on knowledge graphs and DLT demonstrates that multicenter collaboration and secure data exchange can improve clinical decision support; however, such technologies are not yet integrated into anomaly-detection frameworks. Overall, there is a clear gap at the intersection of AI-based anomaly detection, cross-provider interoperability and organisational readiness.

The present study seeks to fill this gap by combining a semi-systematic scoping synthesis with a simulation of cross-provider audit logs. It draws on the readiness frameworks discussed above to develop a four-pillar checklist (governance, infrastructure/interoperability, workforce and AI integration) and pairs this with a pragmatic anomaly-detection evaluation using interpretable rules and Isolation Forest. By addressing both the technical and organisational dimensions, the study aims to contribute a staged implementation pathway for real-world adoption of AI-based anomaly detection across multiple health-care providers.

## 3. Methodology

### 3.1. Introduction

The overall purpose of this study is twofold: to identify the organisational and digital capabilities that healthcare organisations require to adopt cross-provider AI-based anomaly-detection systems (RQ1) and to evaluate the effectiveness and interpretability of lightweight anomaly-detection models when contextual audit features are available (RQ2). A mixed-methods design was adopted to address these goals. A semi-systematic scoping review mapped existing literature on organisational readiness and AI adoption in healthcare, while a rapid review summarised recent anomaly-detection techniques for electronic health record (EHR) audit logs. These review components informed the development of a simulation experiment that emulated cross-provider audit logs to benchmark detection approaches. Scoping reviews are appropriate when the objective is to "map the extent, range and nature of the literature and identify gaps" (Mak & Thomas, 2022), and they employ a systematic and iterative search strategy coupled with clearly defined inclusion criteria (Mak & Thomas, 2022). The simulation approach complements the review by providing empirical evidence on detection performance using synthetic data, a necessary compromise when real cross-provider audit logs are unavailable (Yeng et al., 2021). Together, these methods allow the study to answer both adoption- and effectiveness-focused research questions.

### 3.2. Research design and development

#### 3.2.1. Semi-systematic scoping review and rapid review

The scoping review followed the methodological guidance outlined by Mak and Thomas (2022). After formulating RQ1, input from domain experts and consultation with a health sciences librarian informed the development of the search strategy. The databases searched included Scopus, Web of Science, PubMed, and IEEE Xplore. Search strings combined synonyms for *EHRs*, *audit logs*, *anomaly detection*, *interoperability*, *digital readiness*, and *AI adoption*, using Boolean operators and medical subject headings. Inclusion criteria limited results to peer-reviewed journal or conference papers published between 2011 and 2025, in English, and focused on healthcare contexts. Eligible studies addressed policy, data privacy, anomaly-detection algorithms, digital-readiness frameworks, or AI-adoption barriers, while editorials, commentaries, and non-healthcare studies were excluded. Duplicates were removed using EndNote, and titles and abstracts were screened manually in Microsoft Excel. Full-text screening followed, with exclusion reasons documented. A calibration check on an initial subset of papers was conducted to ensure consistency and alignment with the inclusion criteria (Mak & Thomas, 2022).

In parallel, a rapid evidence review addressed RQ2 by summarising anomaly-detection algorithm families, input features, and evaluation methods. The review emphasised unsupervised, semi-supervised, and deep-learning approaches applied to EHR audit logs, drawing from sources like Tabassum et al. (2024) and Niu et al. (2024). Extracted data encompassed dataset type (real vs. synthetic), feature engineering techniques, algorithm families (e.g., Isolation Forest, Local Outlier Factor, autoencoders, BERT-based models), evaluation metrics, and performance outcomes. This review complemented the scoping review by pinpointing interpretable algorithms suitable for resource-constrained healthcare settings.

#### 3.2.2. Synthetic audit-log simulation

To empirically evaluate detection performance in a cross-provider context, a synthetic audit-log simulation was developed. Real operational logs are seldom shareable because of privacy and liability constraints; reviews of security-practice studies frequently report reliance on synthetic or strongly de-identified data (Yeng et al., 2021). Building on the conceptual workflow in Yeng et al., the simulation replicated typical EHR-access processes from data sources through preprocessing, feature extraction, and classification.

A hypothetical health information exchange was specified with eight providers (public/private hospitals and clinics) sharing a common EHR. Each provider included multiple user roles (physicians, nurses, administrative staff), and each patient was linked to a primary provider. The base log recorded who accessed which patient record, at which provider, for which event type (view/modify/discharge), and at what time.

Synthetic records were generated in Python (pandas, NumPy), using distributions informed by published EHR-use studies. For each session the generator assigned a user role, patient, provider, session start/end times, and event types. Key contextual variables were:

- **Days since discharge**: time since the most recent discharge; accesses beyond a threshold (e.g., >14 days) were marked suspicious.
- **Provider mismatch**: whether the accessing provider differed from the patient's primary provider; legitimate for referrals but potentially risky if repeated or uncoordinated.
- **Session duration**: unusually long/short sessions flagged as potential misuse.
- **Hour of day / day of week**: captures off-hours activity often associated with misuse (Yeng et al., 2021).
- **Recent access frequency**: repeated access by the same user–patient pair within a rolling 24-hour window as a snooping signal.

Anomalies were injected using patterns described in prior misuse-detection studies (Yeng et al., 2021) and aligned with explanation-based auditing logic (Fabbri & LeFevre, **2011**): (1) cross-provider access without a documented referral; (2) post-discharge access beyond a defined interval; (3) off-hours access by roles typically restricted to day shifts; (4) extreme session durations; and (5) rapid repeated access to the same record by the same user.

Two datasets were produced: a **refined** set (~500 sessions; 99 injected anomalies) emphasising core, interpretable features, and a **complex** set (~1,000 sessions; 200 anomalies) with additional noise variables (e.g., shift type, department, multi-patient sessions). Each dataset was split 80%/20% into training and test partitions for model evaluation. Two approaches were compared: (i) a transparent rule-based system parameterised on the contextual features above, and (ii) Isolation Forest as an unsupervised baseline consistent with prior health-access anomaly work (Tabassum et al., 2024).

### 3.3. Data collection and extraction

#### 3.3.1. Study selection

The review followed the PRISMA-ScR checklist for scoping reviews (Tricco et al., 2018) and reports study flow using the PRISMA 2020 format (Page et al., 2021). Database searches across PubMed, Scopus, Web of Science, IEEE Xplore, ACM Digital Library, and arXiv returned 146 records; manual hand searching and backward/forward snowballing added 7 records (total = 153). After de-duplication in EndNote, 128 records remained and were screened by title/abstract in Microsoft Excel; 88 were excluded. Full texts were retrieved for 40 articles; 25 were excluded with documented reasons: not focused on security/privacy or anomaly detection (n = 7); no cross-provider or interoperability context (n = 6); conceptual/opinion with no empirical evaluation (n = 5); no clear AI model or evaluation (n = 3); duplicate/superseded (n = 2); full text inaccessible (n = 2). Fifteen studies were included in the qualitative synthesis.

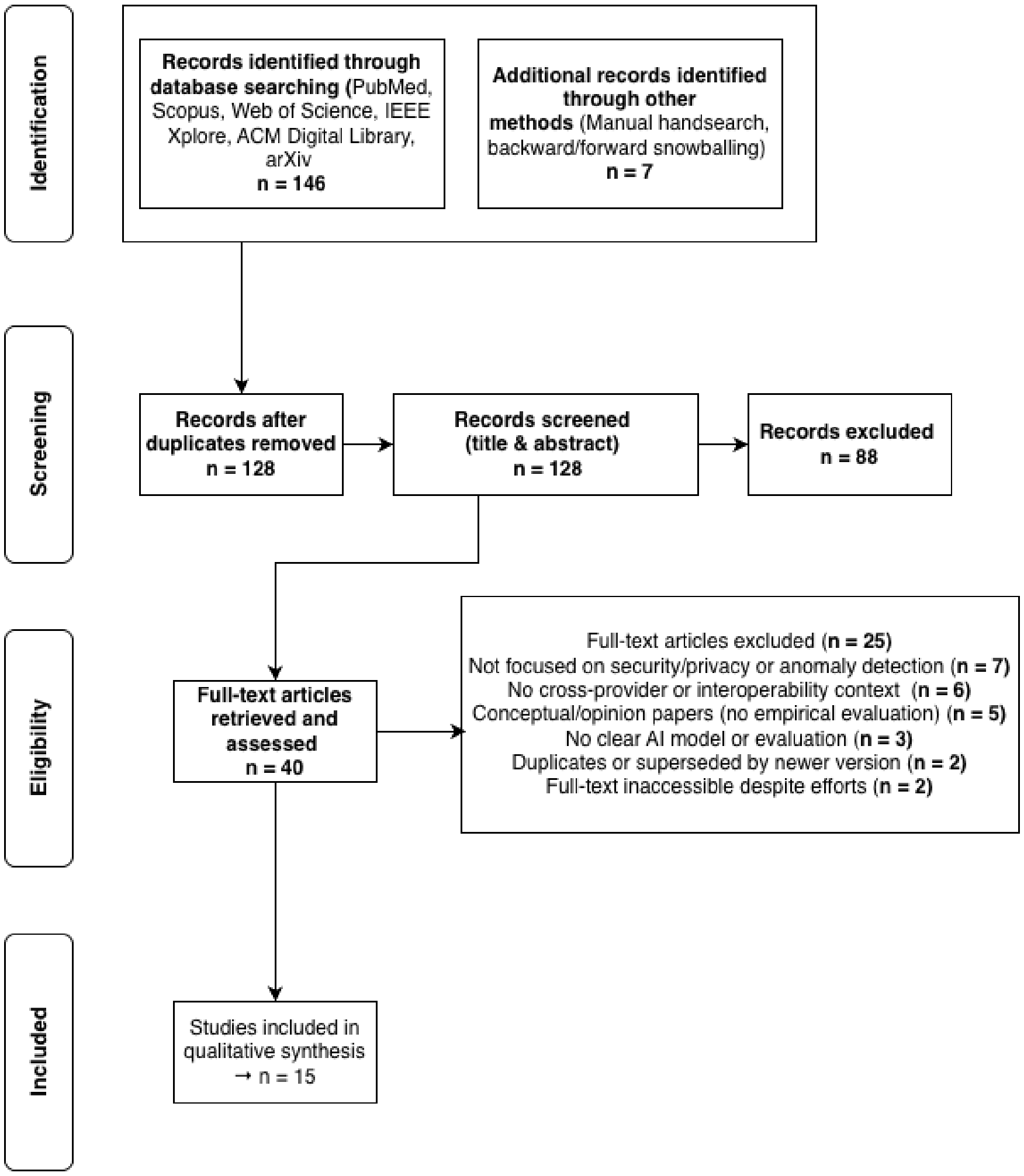

**Figure 1. PRISMA flow diagram for study selection in the scoping review**

### 3.3.2. Data extraction and synthesis

Data extraction was conducted by a single reviewer using a predefined Excel template. Core fields captured author, year/venue, study type, setting, focal topic (audit-log analytics, digital readiness, AI adoption), methods, findings, and reported barriers/facilitators. For anomaly-detection studies, additional fields recorded dataset characteristics (real vs synthetic; single-site vs multi-provider), contextual features (e.g., provider mismatch, time of day), algorithm family, any explainability method, and performance metrics (precision, recall, F1). The template was piloted on five articles to calibrate definitions; an audit trail documented inclusion/exclusion decisions. A best-fit framework synthesis was applied: deductive codes (governance; infrastructure/interoperability; workforce capability; AI integration) were seeded from established frameworks and complemented with inductive codes emerging from the studies, then mapped to the four-pillar readiness checklist.

### 3.3.3. Rapid-review data collection

For the rapid review, summary data were extracted from each study's methods/results: log/data type (audit log, network log, claims), data status (real vs synthetic), feature-engineering strategies, algorithm family (e.g., Isolation Forest, Local Outlier Factor, autoencoders, BERT-based models), reported interpretability technique (e.g., rules, SHAP, explanation-based auditing), and performance metrics. These data informed algorithm selection and feature design for the simulation.

#### 3.3.4. Simulation data collection

The simulator produced structured audit-log datasets exported to comma-separated values (CSV). Each row represented a user–patient access session and contained: anonymised user ID, provider ID, patient ID, event type, event timestamp, session duration, discharge timestamp, primary provider, derived fields (days since discharge, hour of day, day of week), and a binary anomaly label. The basic dataset included only these variables to support interpretability. The complex dataset added contextual noise fields (e.g., department, shift type) to test model robustness under less controlled conditions. Datasets were saved locally and also provided in spreadsheet format for analysis. Reproducibility was supported by version-controlled Python scripts (fixed random seed, documented preprocessing and feature derivations) and a data dictionary describing variable definitions and coding.

### 3.4. Data analysis techniques and tools

#### 3.4.1. Scoping review synthesis

Descriptive statistics summarised publication counts by year, country, and health-care setting. A narrative synthesis then mapped reported **barriers** and **facilitators** to readiness domains using a best-fit framework approach. Deductive codes drew on UTAUT constructs—performance expectancy, effort expectancy, social influence, and facilitating conditions—and organisation-level readiness criteria adapted from recent health-system studies (e.g., governance and vision, infrastructure, agility, workforce capability) (Gardner et al., 2023). Inductive codes were added where concepts did not fit the seed frames. The synthesis yielded a **four-pillar, 10-item readiness checklist** aligned to **Governance**, **Infrastructure & Interoperability**, **Workforce**, and **AI Integration**, with each item linked to concrete evidence artefacts (e.g., standardised audit fields, RBAC and log-retention SLAs, escalation playbooks). Heatmaps of coded frequencies (readiness themes and barrier themes) provide a cross-study view of emphasis and gaps (see Figures 2–3).

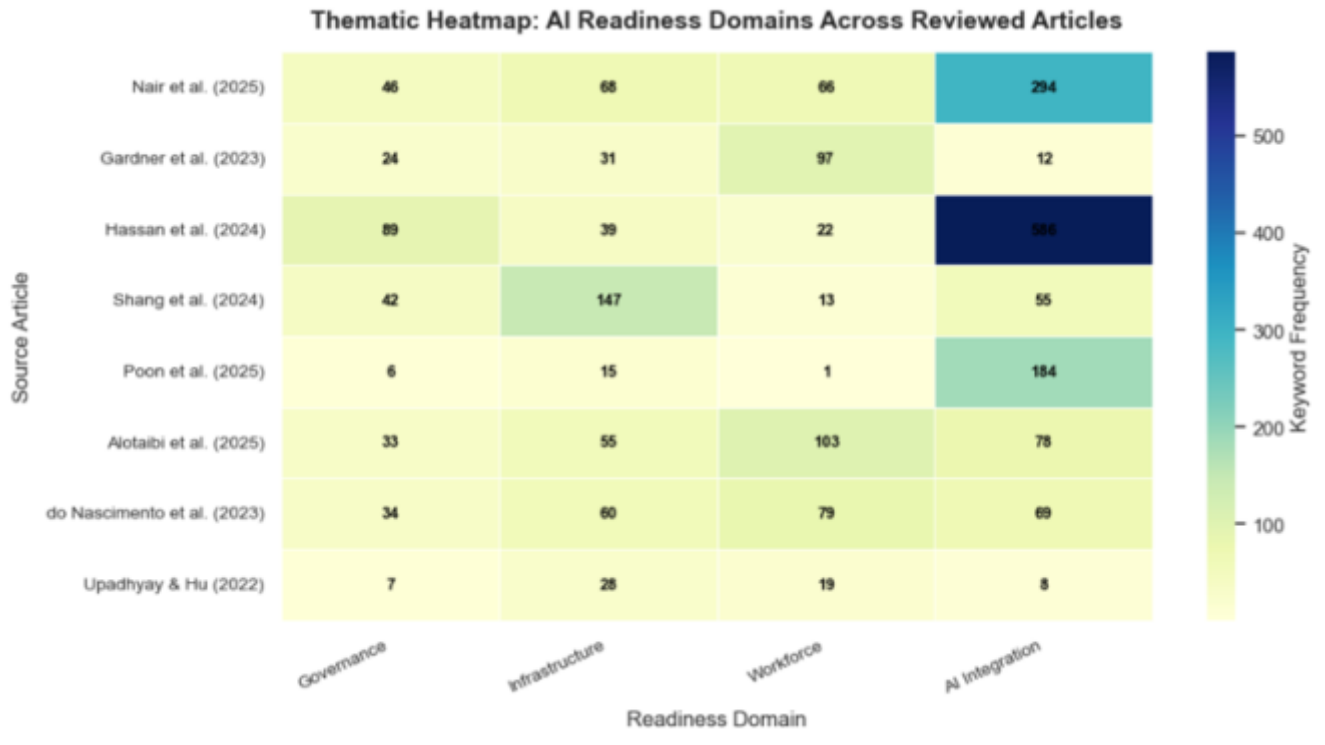

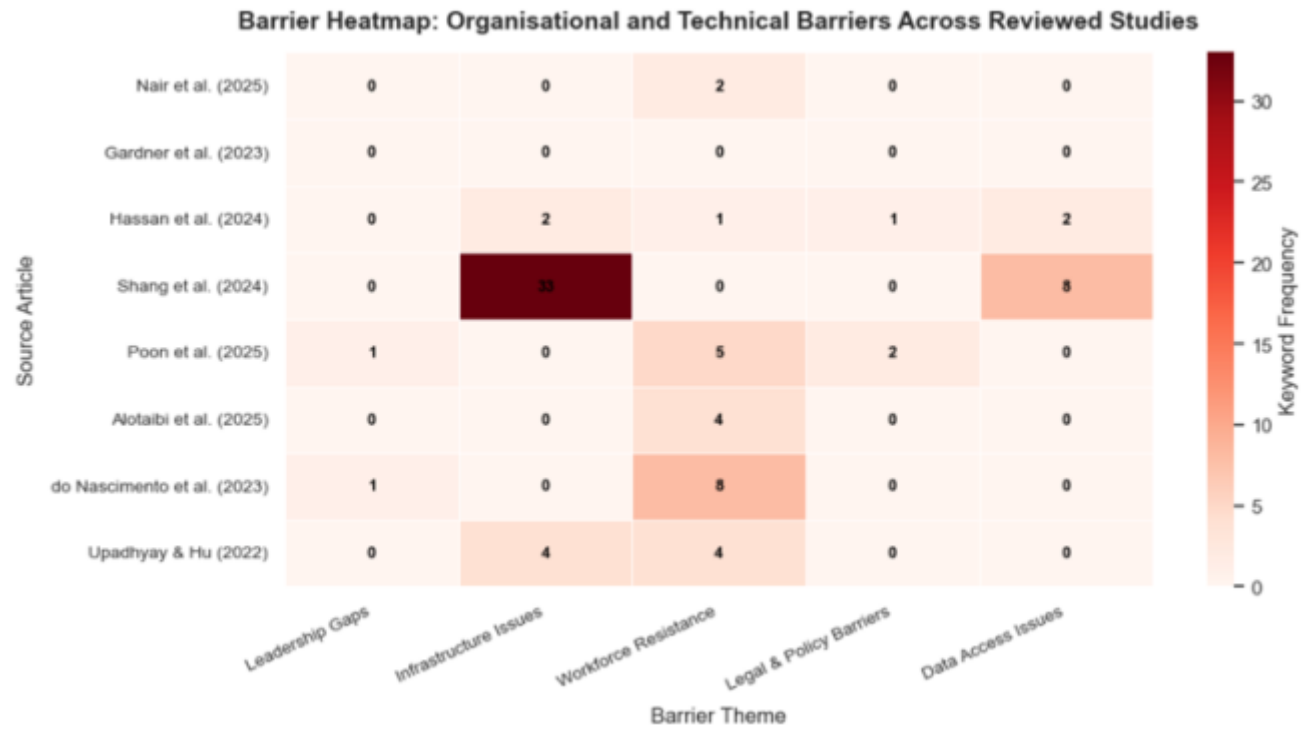

**Figure 2**

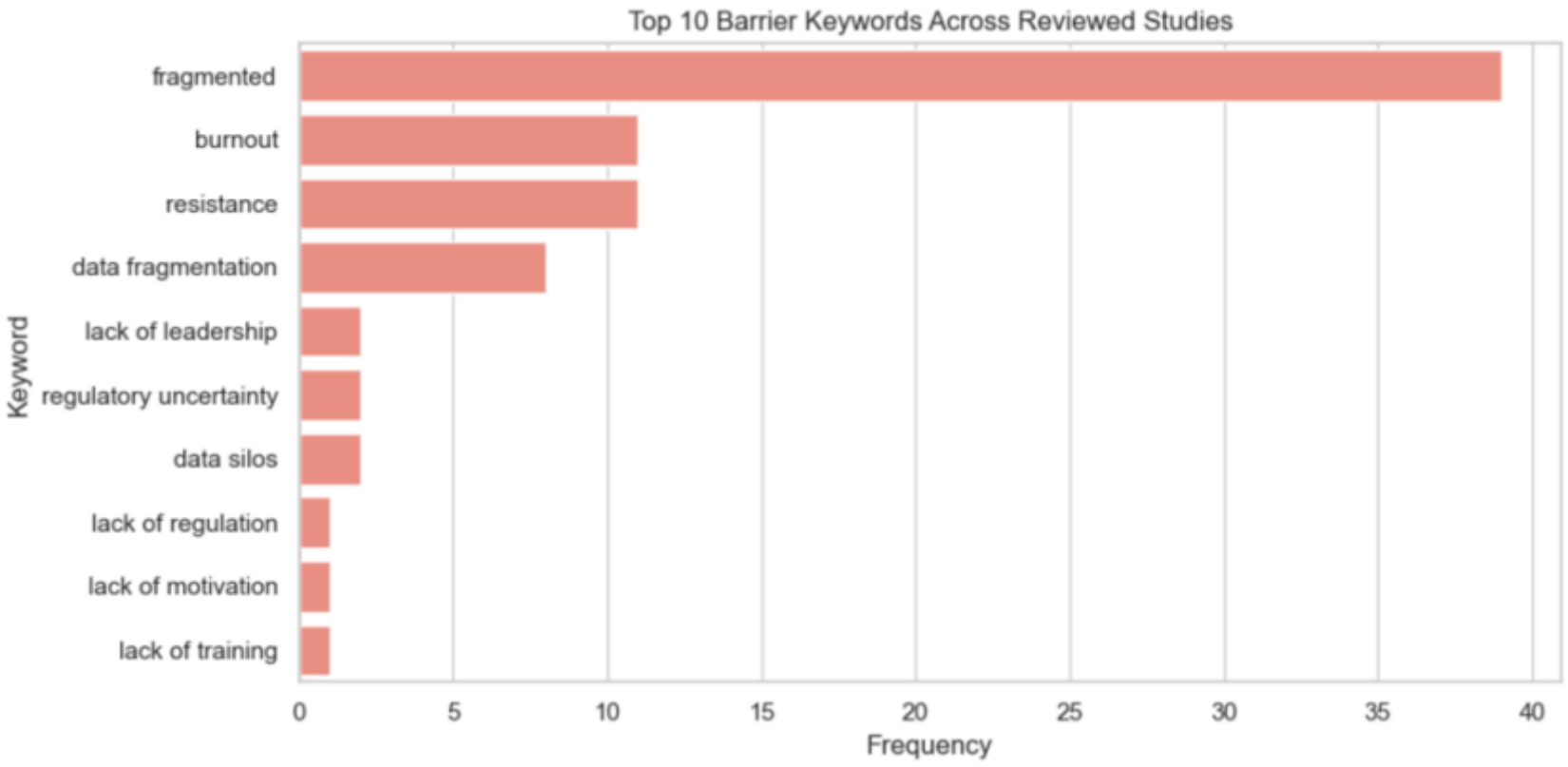

**Figure 3**

#### 3.4.2. Simulation modelling and evaluation

Two anomaly-detection approaches were benchmarked on synthetic cross-provider audit logs:

Rule-based classifier. A transparent rule set instantiated the anomaly definitions: e.g., flag `days_since_discharge` > 14; flag `provider_mismatch` without a referral marker; flag off-hours access by day-shift roles; flag extreme session duration; flag rapid repeats for the same user–patient pair. Rules emitted binary indicators; records were classified anomalous if any indicator was positive. Thresholds were tuned on the training set to balance recall and false positives.

Isolation Forest. Isolation Forest was selected as an unsupervised baseline because prior health-security studies report competitive performance on limited, heterogeneous data and do not require labels (Tabassum et al., 2024). The scikit-learn implementation used 100 trees, `max_samples`="auto", and contamination set to the known anomaly rate; features were standardised. To support adoption, interpretability was foregrounded: rules are inherently interpretable, and for Isolation Forest, SHAP (Shapley Additive Explanations) was used to attribute anomaly scores to features and to visualise global importance and interactions (e.g., `provider mismatch` × `off-hours`) (Niu et al., 2024).

Performance metrics. Precision, recall, F1-score, and ROC-AUC were computed for both classifiers. The refined and complex datasets were evaluated separately to assess robustness under additional contextual noise. To gauge variability, training/evaluation was repeated across five random seeds and the mean ± SD reported. (See metric tables, confusion matrices, agreement plots, and SHAP summaries in the results/appendix.)

### 3.5. Validation and limitations

Scoping review. Rigour was supported by a written protocol, EndNote de-duplication, Excel-based screening, a calibration pass on an initial subset, and an auditable record of inclusion/exclusion decisions.

The review is limited to English-language, peer-reviewed literature to 2025 and emphasises breadth over critical appraisal, consistent with scoping aims (Mak & Thomas, 2022). Some relevant grey literature may have been missed, and coding depended on a single reviewer.

Simulation. Synthetic logs enabled cross-provider evaluation without risking patient privacy, consistent with prior security-practice work that frequently relies on synthetic or strongly de-identified data (Yeng et al., 2021). Nevertheless, synthetic sessions cannot capture all workflow nuances; anomaly patterns were templated from published heuristics and explanation-based auditing logic (Fabbri & LeFevre, 2011), so real misuse behaviours may differ. The sample size was modest, and only a rules baseline and Isolation Forest were benchmarked. Deep models (e.g., EHR-BERT) were not implemented due to computational constraints, although prior work suggests potential accuracy gains at the cost of transparency and resources (Niu et al., 2024). Findings should be validated on real multi-site audit logs, with expanded method families (e.g., LOF, OC-SVM, autoencoders, sequence/ensemble models) and prospective assessment of reviewer workload and trust.

## 4. Results & Discussions

### 4.1. Results

#### 4.1.1. Readiness checklist

The semi-systematic review yielded 15 papers that addressed adoption or implementation of anomaly-detection systems, digital readiness or AI governance. Using thematic analysis, findings were synthesised into a four-pillar readiness checklist: Governance, Infrastructure/Interoperability, Workforce and AI integration.

| Pillar | Objective | Core capabilities | Key organisation-agnostic controls (reusable) | Evidence to produce (auditable artefacts) | Risks if absent |
|---|---|---|---|---|---|
| **Governance** | Establish accountable, lawful, and transparent oversight of AI-enabled audit monitoring. | Executive sponsorship; policy ownership; escalation accountability; privacy & compliance assurance. | • Named risk owner for alert thresholds & SLAs<br>• Policy for cross-provider audit-log sharing and investigation<br>• RACI for triage/escalation<br>• Data-protection impact assessment (DPIA) & data-sharing agreements (DSA/DPA). | Board/committee minutes assigning ownership; approved threshold/SLA policy; escalation playbook; DPIA excerpts; signed DSA/DPA; change-control log for threshold revisions. | Unclear accountability; unlawful data flows; inconsistent responses to alerts; regulatory exposure. |
| **Infrastructure & Interoperability** | Provide reliable, secure, and standardised data flows across sites. | Common audit schema; identity/role quality; secure data exchange; pipeline observability. | • Minimum core fields (action, timestamp, user, patient, role, location, provider)<br>• Identity resolution across sites; RBAC metadata alignment<br>• Secure ETL/API with completeness/latency monitoring<br>• Versioned data dictionary. | Schema + dictionary; IAM reconciliation reports; role mapping rates; pipeline run logs & dashboards; data-quality checks (missingness, timeliness); API/ETL design docs. | Fragmented logs; identity mismatches; data loss or delay; low signal-to-noise for detectors. |
| **Workforce** | Ensure people can interpret alerts and act proportionately. | Designated reviewers; protected time; training; clinician communication and feedback channels. | • Named reviewer roles with duty roster<br>• Curriculum for audit-log interpretation & tool use<br>• Communication plan to clinicians (what alerts mean; how to contest) • Feedback loop to improve rules/models. | Training records; competency checks; standard operating procedures (SOPs); comms artefacts (briefings, FAQs); ticketing/feedback logs linked to corrective actions. | Alert fatigue; mistrust of AI; inconsistent dispositions; latent safety/privacy risks. |
| **AI Integration** | Treat the model as a managed, explainable, and continually improved component. | Explainability (global/local); performance & drift monitoring; retraining/recalibration; documentation. | • Reviewer-facing explanations (e.g., SHAP at case level + global summaries)<br>• Metric dashboards (precision/recall, alert volume)<br>• Data/model drift monitors with triggers<br>• Model cards covering data, assumptions, and limits. | Example alert reports with explanations; model-performance dashboards; drift reports; retraining/change logs; model card / validation memo. | Opaque decisions; silent performance decay; loss of trust; inability to justify actions to regulators. |

**Table 1. Four-pillar readiness framework for cross-provider audit-log anomaly detection**

| ID | Pillar | Checklist item (what to have in place) | Why it matters | Accountable owner | Evidence / measurable indicator | Minimum standard | Review cadence |
|---|---|---|---|---|---|---|---|
| G1 | Governance | Named senior owner for audit-log monitoring and thresholds (RACI defined) | Establishes accountability for risk and resources | CIO/CCO (privacy) | RACI chart; appointment minute; role description | One executive and one operational delegate named | Annual or on role change |
| G2 | | Policy for cross-provider audit-log sharing & investigation | Lawful basis and consistent handling across sites | Privacy office | Approved policy; DPIA; DSAs/DPAs | Policy approved and communicated; DPIA completed | Annual |
| G3 | | Escalation playbook (triage → investigation → outcome) | Reduces variance and delays in responses | Compliance lead | SOP/playbook; example tickets with timestamps | ≤2 business days from alert to triage; time-to-closure target defined | Quarterly KPI review |
| G4 | | Threshold/change control process for rules/models | Prevents untracked drift in alerting behaviour | Data governance board | Change log; approval records; versioned configs | All changes ticketed and approved before deployment | Per change |
| G5 | | Auditability of decisions (who reviewed, rationale, outcome) | Supports accountability to patients/regulators | Audit committee | Case logs with reviewer ID and rationale; retention policy | ≥95% alerts with documented disposition | Monthly spot check |
| I1 | Infrastructure & Interoperability | Common audit schema across sites (user, patient, role, provider, event, timestamp, location, device, outcome) | Enables consistent features and joins | Enterprise architect | Data dictionary; schema conformance report | ≥98% fields present and typed per spec | Monthly |
| I2 | | Identity resolution & role mapping across providers | Avoids false signals and missed links | IAM lead | Match rates; role mapping table; exceptions log | ≥97% deterministic/probabilistic matches; ≤3% unmapped roles | Monthly |
| I3 | | Secure ETL/API pipelines with observability (completeness/latency) | Ensures timely, trustworthy data | Data engineering | Pipeline dashboards; alerting rules; SLA | ≥99% record completeness; <24h | Daily/weekly |
| | | | | | | latency to monitoring store | |
| I4 | | Data quality checks (missingness, duplicates, time skew) | Protects model performance & explanations | Data stewardship | DQ report; automated tests in CI/CD | All checks green or exceptions waived | Each release |
| I5 | | Role-based access control (RBAC) & least-privilege for monitoring tools | Reduces insider risk | Security lead | Access reviews; privilege matrix | Quarterly access recertification; SoD conflicts resolved | Quarterly |
| W1 | Workforce | Designated reviewer function with protected time | Ensures alerts are actually handled | Operations manager | Roster; capacity plan | Coverage ≥5 days/week; backlog ≤2 weeks | Weekly |
| W2 | | Training on audit-log interpretation, SHAP-based explanations, and bias awareness | Builds consistent dispositions and trust | Training lead | Curriculum; attendance; competency check | ≥90% reviewers certified; ≥80% pass on assessment | Semi-annual |
| W3 | | Clinician communication plan (what triggers alerts; how to contest) | Minimises friction and improves data quality | CMIO/Clinical liaison | Briefings; FAQs; intranet page; feedback stats | ≥2 comms per year; ticket response SLA defined | Semi-annual |
| W4 | | Feedback loop from reviewers/clinicians to improve rules & features | Converts operational learning into product changes | Product owner | Backlog with tagged feedback; release notes | ≥1 improvement merged per quarter | Quarterly |
| A1 | AI Integration | Model card & validation memo (data, assumptions, limits, fairness) | Transparent documentation for governance | ML lead | Model card; validation results; sign-off | Published before go-live and on each major update | Per release |
| A2 | | Reviewer-facing explanations (global + local SHAP) embedded in UI | Supports defensible decisions & triage | ML lead / UX | Screenshots; example cases; usability test notes | Global feature summary + per-case SHAP available | Each release |
| A3 | | Performance dashboard (precision/recall, alert volume, time-to-disposition) | Tracks value and burden | Analytics lead | Live dashboard; monthly KPI pack | Targets set by governance (e.g., precision ≥0.6, time-to-triage ≤2 days) | Monthly |
| A4 | | Data/model drift monitoring with triggers (recalibration/retraining) | Prevents silent degradation | ML ops | Drift metrics (PSI/KS); trigger thresholds; retrain log | Drift < threshold or retrain executed within 30 days | Monthly |
| A5 | | Safe rollout & rollback (A/B, shadow mode, canary) | Lowers deployment risk | DevOps | Runbooks; deployment history | Rollback tested in last 6 months | Per release |
| A6 | | Post-implementation review (PIR) including false-positive/negative audit | Institutional learning | Governance board | PIR report; actions & owners | Completed within 90 days of go-live and annually | Annual |

**Table 2. Readiness checklist for cross-provider audit-log anomaly detection**

The governance pillar emphasises leadership support, clear policies and a culture of privacy. Studies stressed that effective adoption requires executive sponsorship, clear data-sharing agreements, and accountability structures for investigation and escalation (Gardner et al., 2023). The infrastructure/interoperability pillar covers technical systems – standardising audit fields (e.g., user, provider, patient, event type, timestamp), implementing identity resolution across providers, and ensuring secure data-exchange protocols. Authors argued that the fragmented nature of current EHR ecosystems compromises anomaly detection and called for harmonised metadata, interoperable audit logs and, where appropriate, distributed ledger or knowledge-graph solutions to maintain data integrity (Ferreira et al., 2024). The workforce pillar highlights skills, training and cultural readiness; prior reviews found that ease of use, training and support are critical facilitators, whereas complexity and lack of digital literacy are barriers (Alotaibi et al., 2025). Finally, the AI-integration pillar synthesises findings on algorithm selection, validation and monitoring. Readiness frameworks emphasise the need for continuous evaluation, governance of AI models, and transparency measures such as explainability (Nair et al., 2025).

The resulting checklist operationalises these themes into 10 items (e.g., presence of audit policies, standardised event taxonomies, role-based access controls, escalation playbooks, staff training programmes, routine model audits and recalibration triggers), providing practical guidance for organisations planning cross-provider anomaly detection.

#### 4.1.2. Baseline performance

Two synthetic audit-log datasets were analysed: Refined (≈500 sessions, 99 anomalies) and Complex (≈1,000 sessions, 200 anomalies). Descriptive statistics showed that session durations were right-skewed, access times covered the full 24-hour cycle, and days since discharge followed a long-tailed distribution. Spearman correlations between features and the anomaly label were low ($|\rho| < 0.1$), indicating that non-linear relationships dominate and motivating the use of model-agnostic explainability techniques.

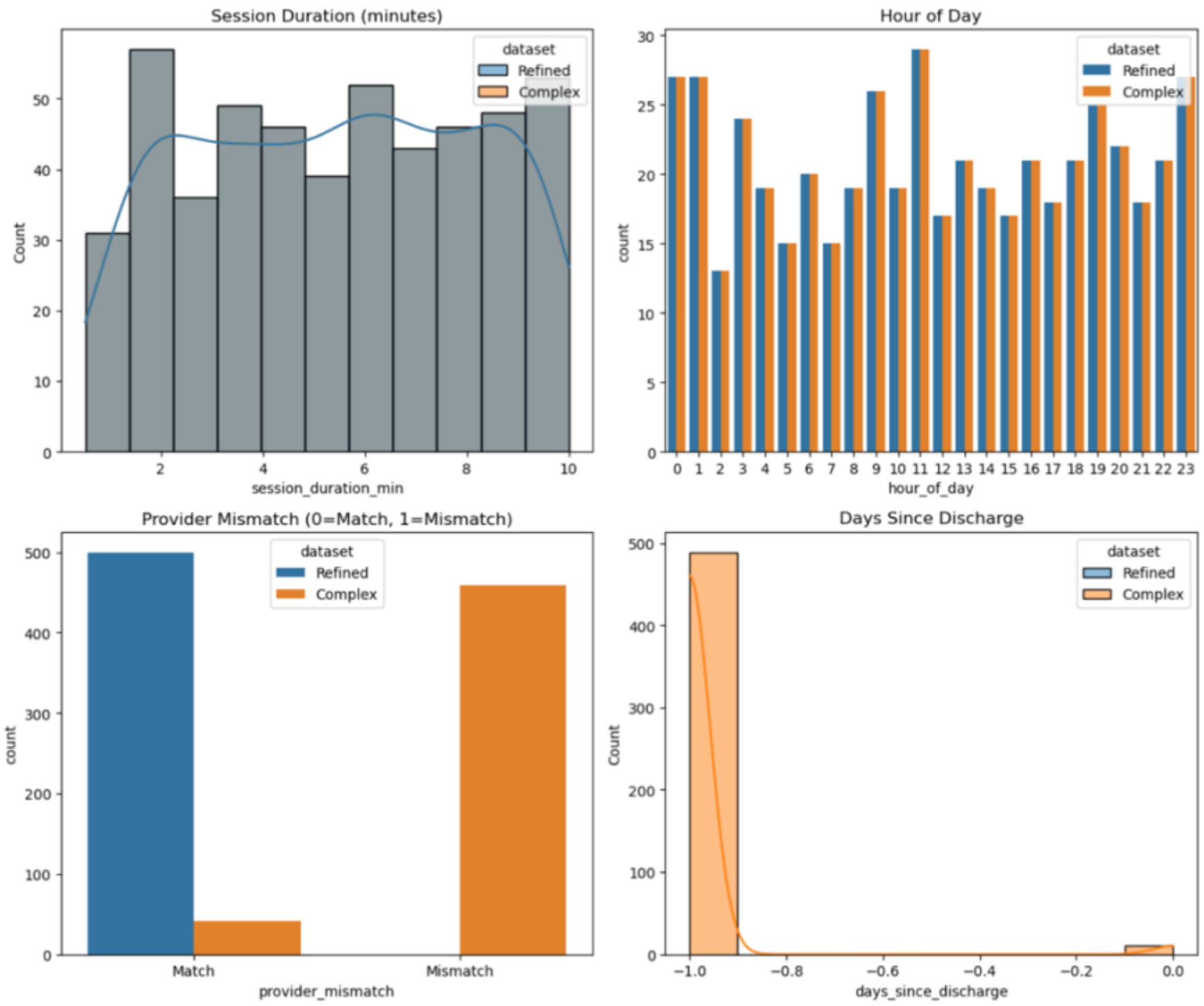

**Figure 4.** Core feature distributions across datasets

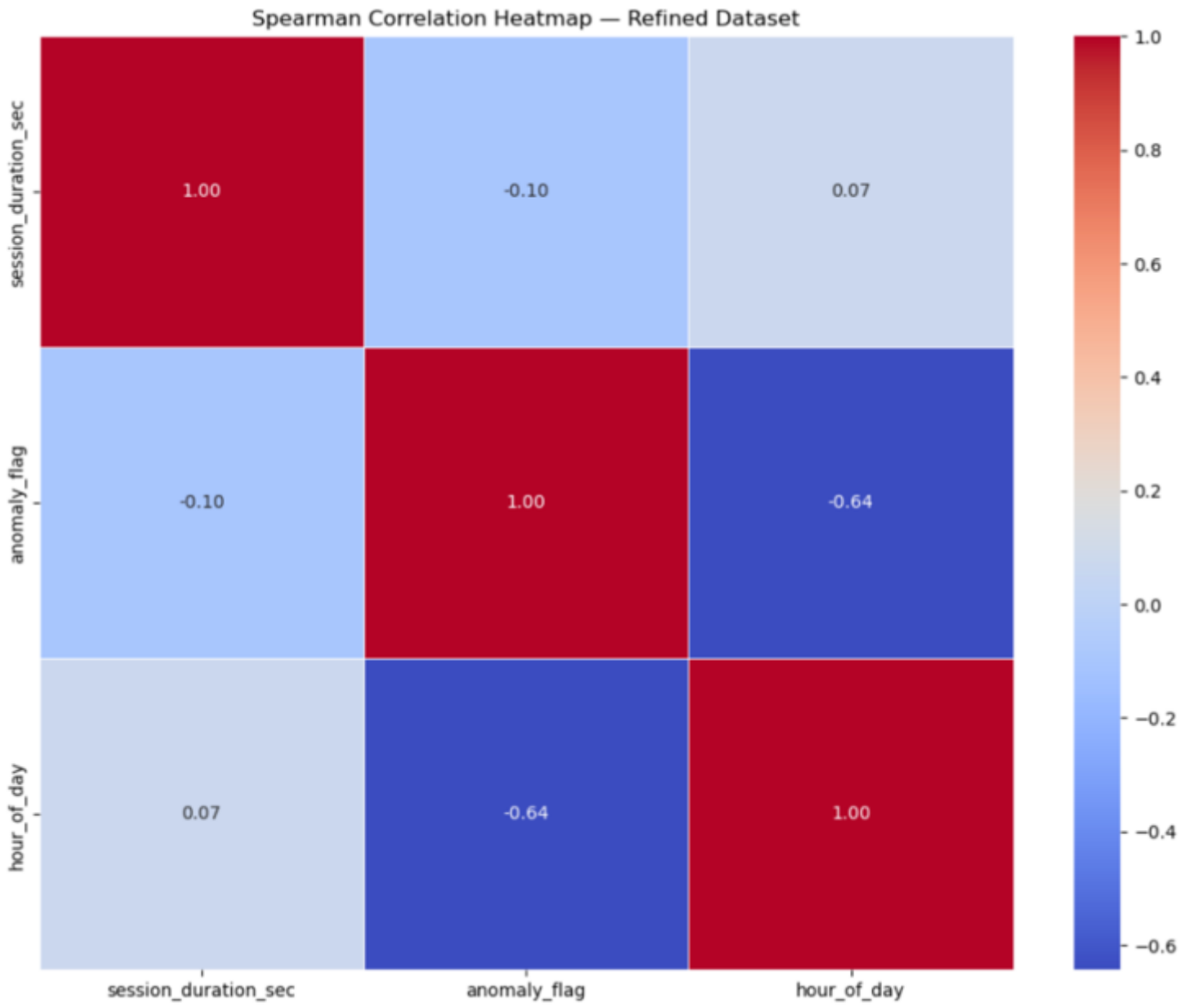

**Figure 5.** Spearman correlation heatmap—Refined dataset.

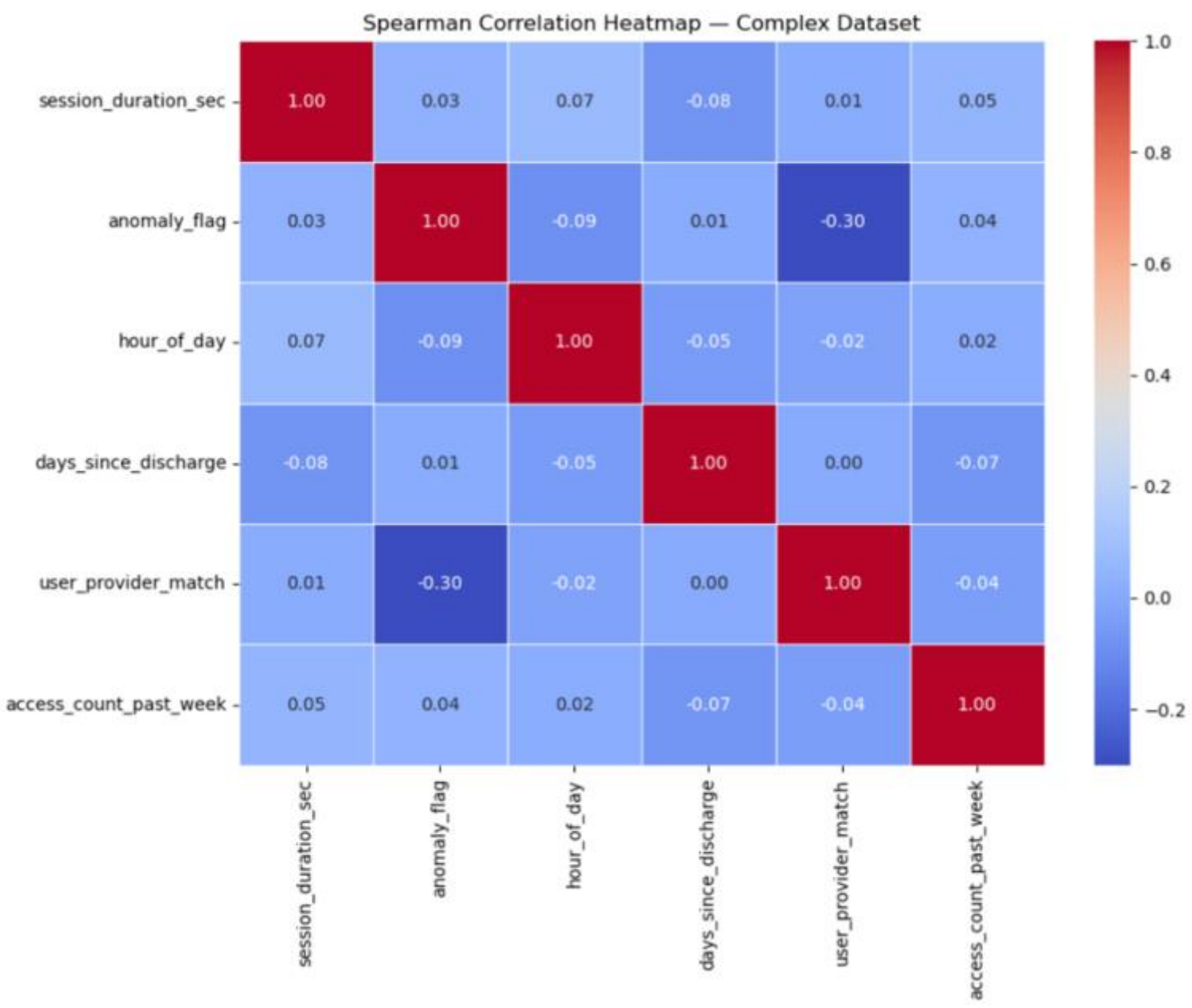

**Figure 6.** Spearman correlation heatmap—Complex dataset.

On the Refined dataset, the rule-based baseline failed to capture any anomalies (recall = 0.00), highlighting the limitation of simple thresholds when provider mismatches are rare. Isolation Forest (IF) achieved modest recall (~0.23) with a low false-positive rate (~0.07 of normals flagged), yielding a precision of roughly 0.77. ROC and precision–recall curves placed IF above the rule-based point but far from ideal sensitivity.

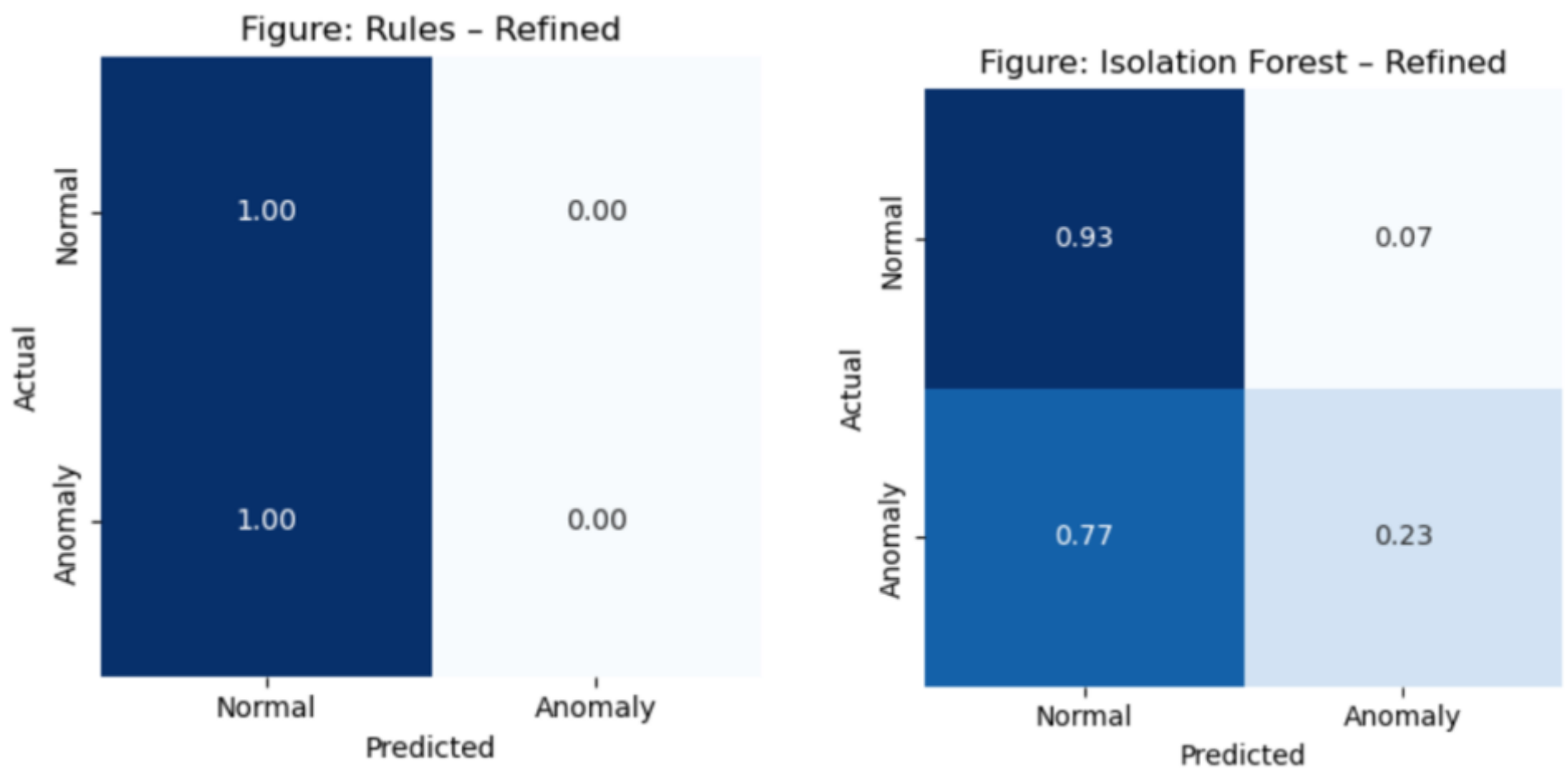

**Figure 7.** Confusion matrices—Rules vs Isolation Forest (Refined).

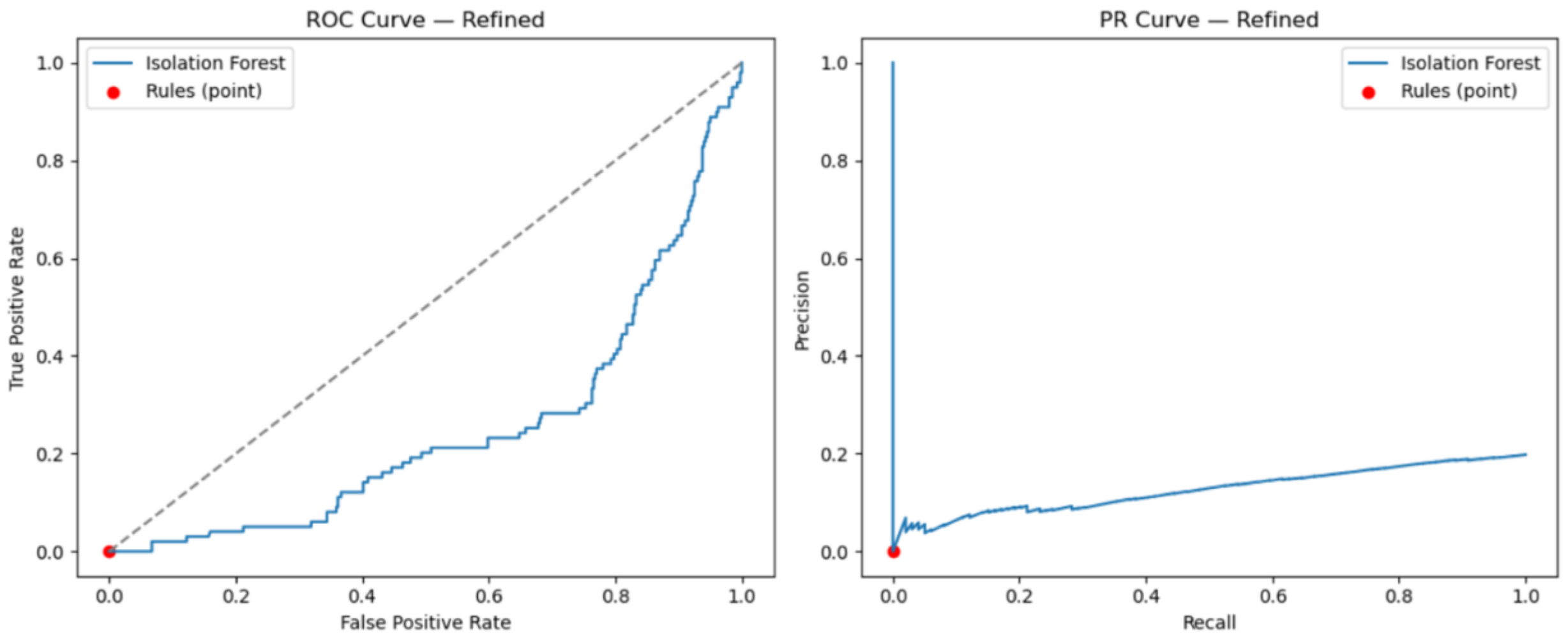

**Figure 8.** ROC and PR curves—Isolation Forest; rules shown as a point.

On the Complex dataset, which included more noise variables and a higher rate of cross-provider accesses, the rule-based system captured roughly half of the injected anomalies (recall ≈ 0.48) at the cost of a noticeable false-positive rate. In contrast, IF exhibited high precision (> 0.9) with very low recall (~0.09), generating only a few alerts. Alert-burden plots showed that IF reduced alert volume by 79 % compared with rules but missed ~91 % of anomalies. These results illustrate a fundamental trade-off: interpretable rules ensure coverage (high recall) but flood reviewers with alerts, whereas IF provides prioritisation (high precision) but risks missing many anomalous sessions.

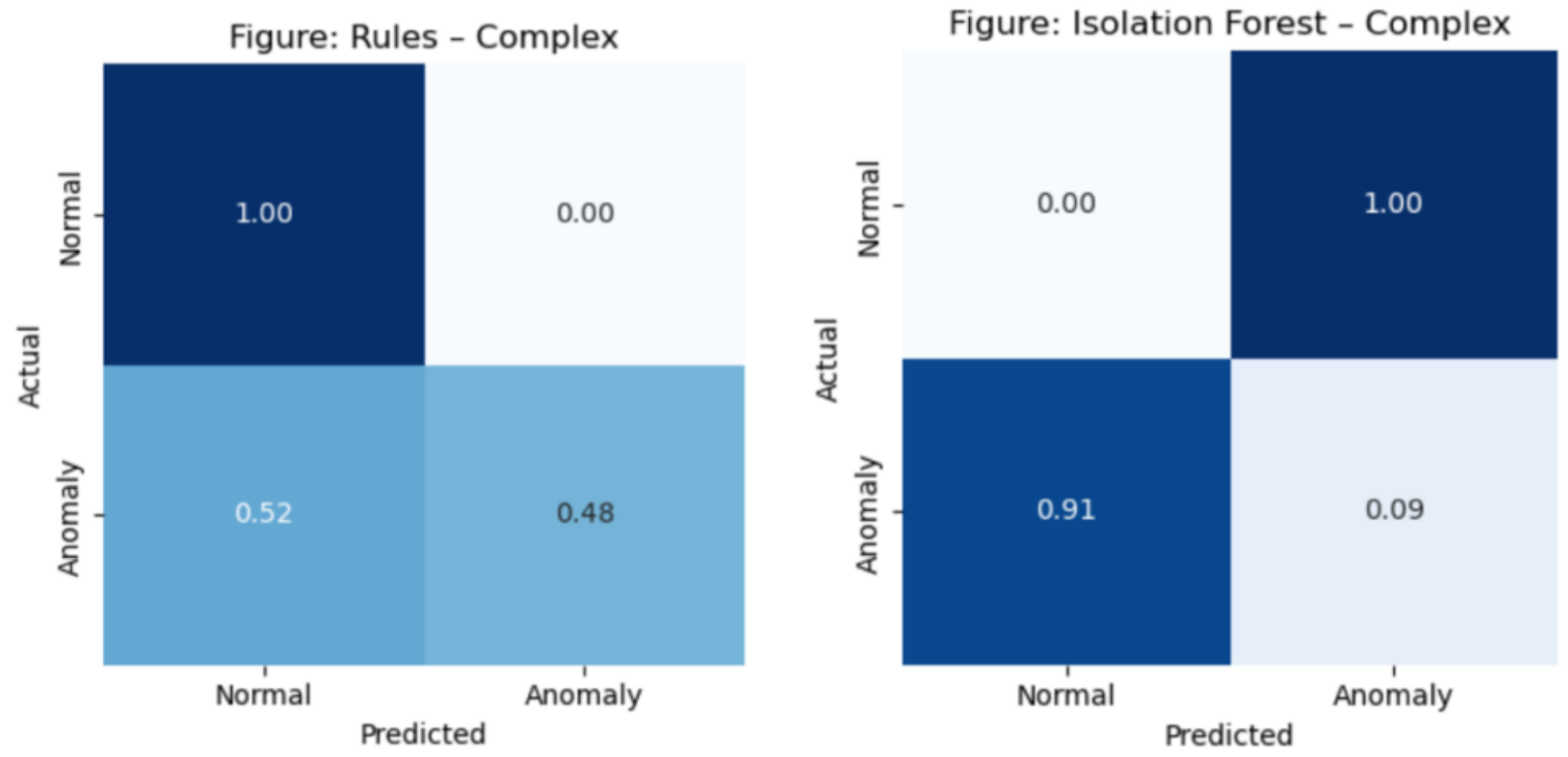

**Figure 9.** Confusion matrices—Rules vs Isolation Forest (Complex).

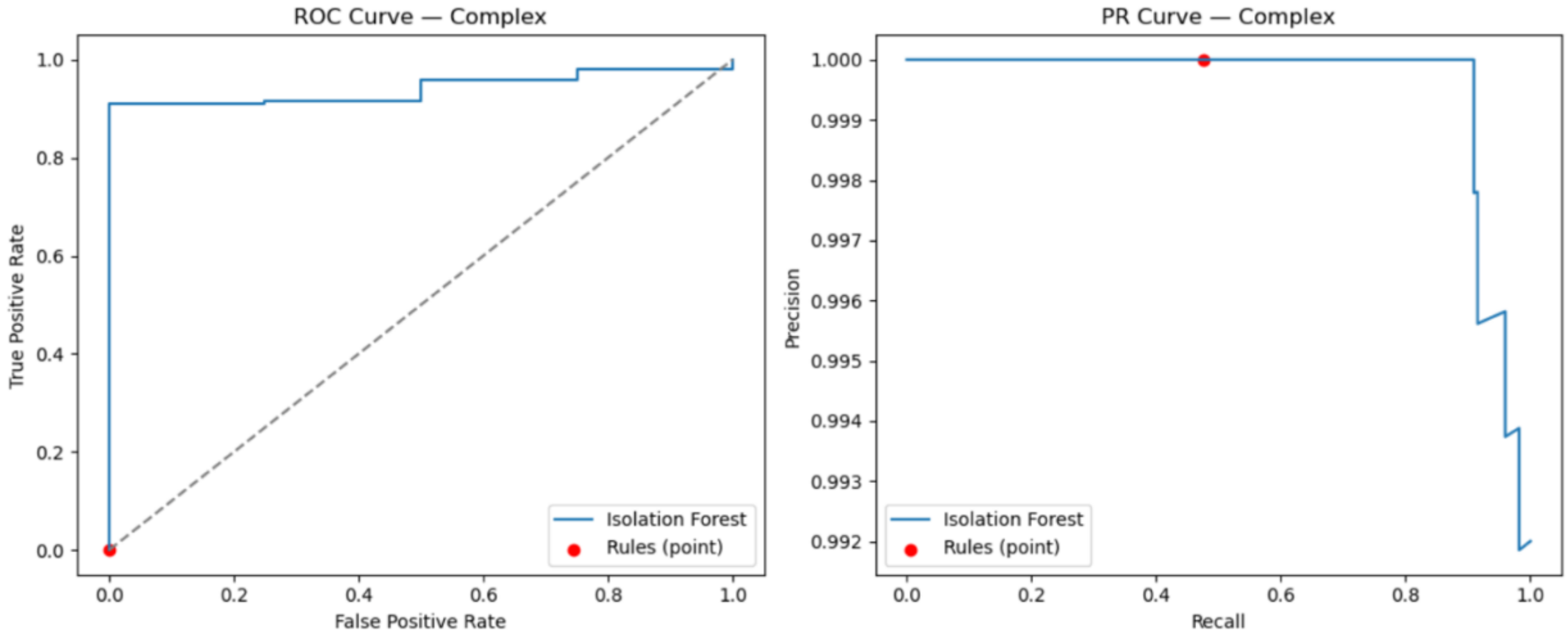

**Figure 10.** ROC and PR curves—Isolation Forest; rules shown as a point.

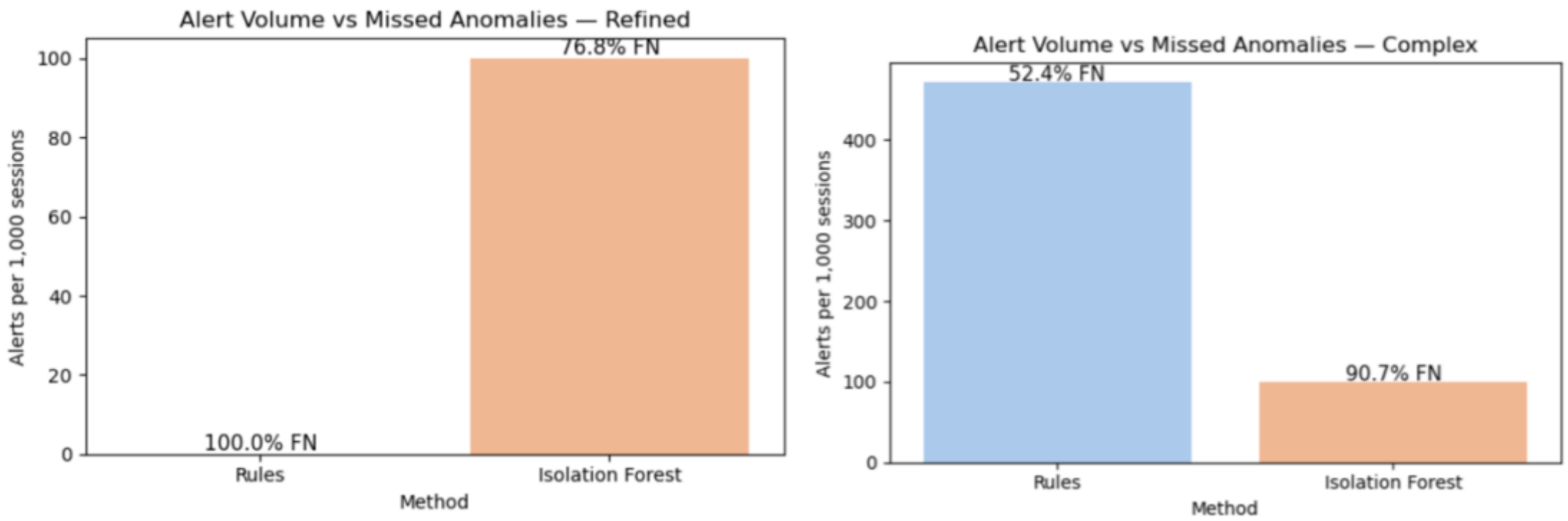

**Figure 11.** Alerts per 1,000 sessions and missed-anomaly rates—Refined and Complex

#### 4.1.3. SHAP explanations for Isolation Forest

To examine why IF prioritised certain records over others and to provide actionable insights for reviewers, Shapley Additive Explanations (SHAP) were computed on the Complex dataset. Global feature importance, illustrated by the mean absolute SHAP values (Figure 12), reveals that provider mismatch is the most influential driver of IF's anomaly scores (mean |SHAP| ≈ 0.6). Hour of day (≈0.38), access_count_past_week (≈0.35), session_duration_sec (≈0.34) and days_since_discharge (≈0.18) follow in descending order of importance. The dominance of provider mismatch underscores that cross-provider accesses are the primary signal used by the model to separate normal from anomalous sessions. This finding aligns with misuse-detection frameworks which highlight cross-provider access without a documented referral as a hallmark of inappropriate behaviour (Yeng et al., 2021).

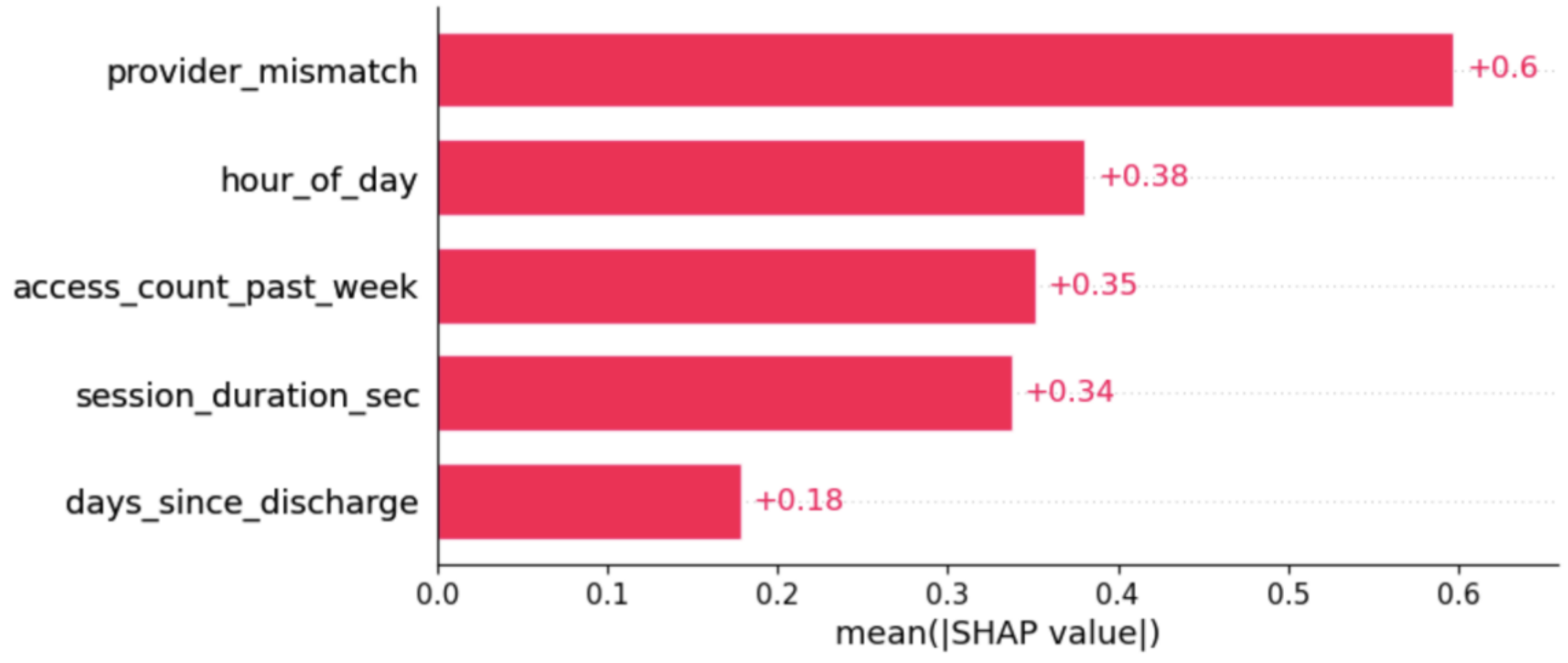

```
Text(0.5, 0.98, 'Global SHAP Importance – Complex Dataset')
```

**Figure 12. Global SHAP Importance – Complex Dataset**

The SHAP summary plot (Figure 13) provides a more granular view: each point represents a session, with colour indicating the feature value. Sessions with provider mismatch = 1 (red points) contribute positively to the anomaly score, whereas mismatch = 0 (blue) have negative SHAP values, pulling predictions towards "normal." Off-hours accesses (hour_of_day values near midnight and early morning, coloured red) correspond to higher SHAP values, reflecting the model's learned association between unusual access times and anomalies. Similarly, high access_count_past_week values push predictions towards anomalous. Longer session_duration_sec values exert a moderate positive influence, whereas extremely short sessions (e.g., quick look-ups) tend to reduce anomaly scores. Days_since_discharge shows a threshold effect: sessions occurring well after discharge (> 14 days) have positive SHAP values, while recent accesses (< 1 day) have negative contributions.

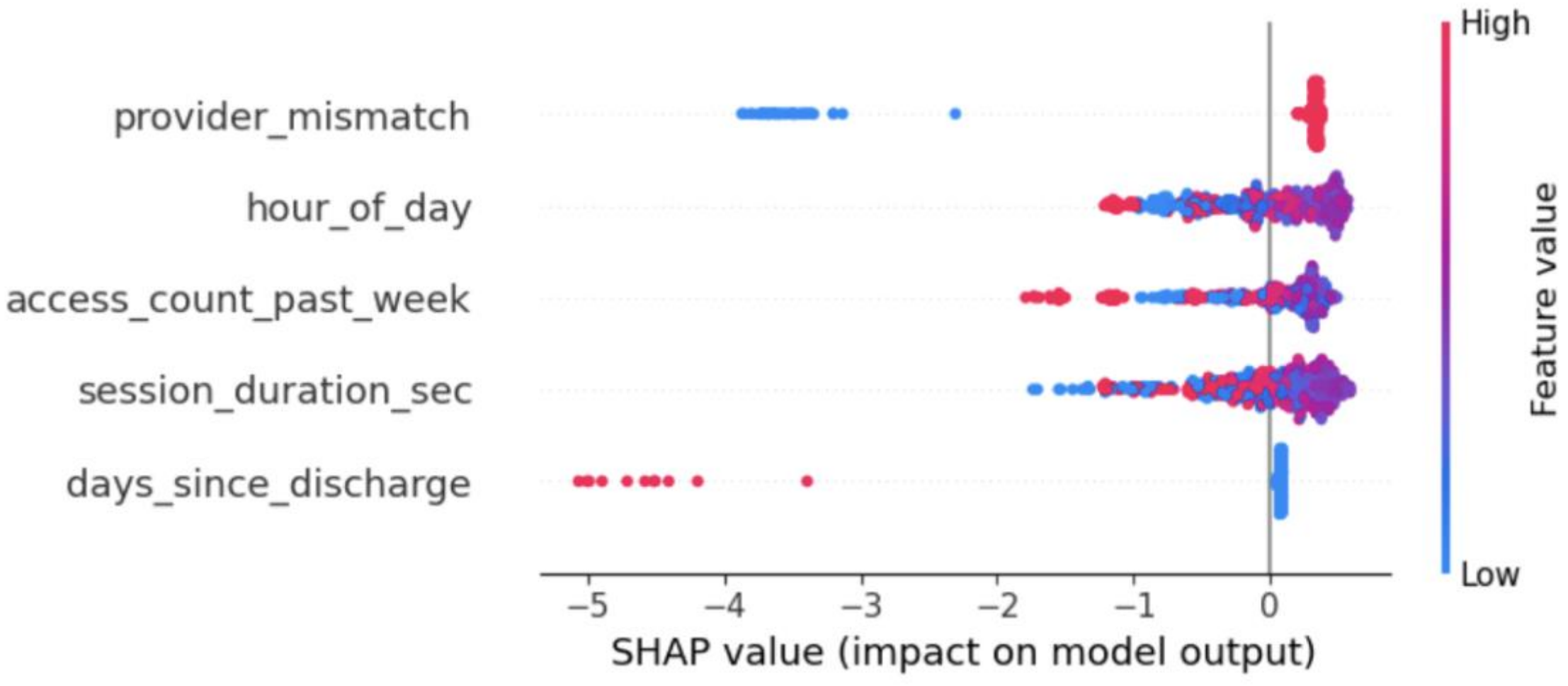

**Figure 13. SHAP summary plot**

Dependence plots illustrate interactions between features. Plotting the SHAP value of provider mismatch against days_since_discharge (Figure 14) shows two distinct clusters: when mismatch = 0, SHAP values cluster around –4, indicating strong negative influence; when mismatch = 1, values rise above +4. Colour gradients representing hour_of_day and access_count_past_week reveal that the highest anomaly scores occur when provider mismatches coincide with off-hours (purple/red) or when the same user–patient pair has numerous recent accesses. This interaction suggests that isolated mismatches may be benign (e.g., a referral), but repeated off-hour mismatches or mismatches after discharge are highly suspicious. Such patterns would be difficult to capture with simple thresholds alone and underscore the value of model-agnostic explanations.

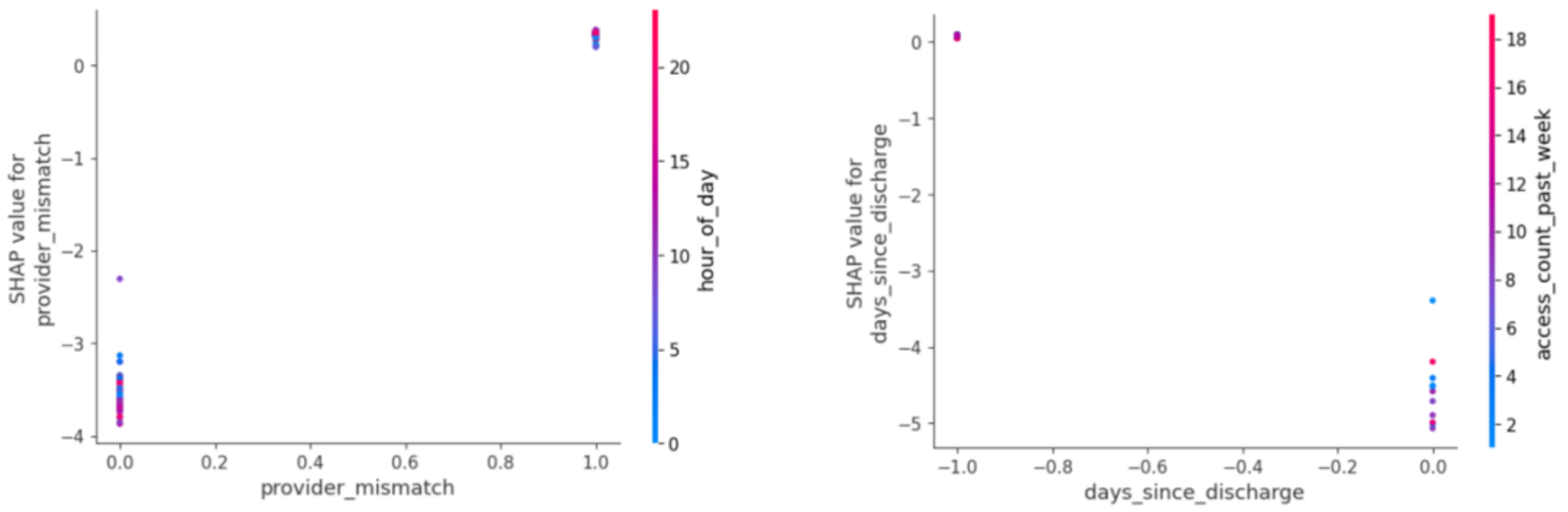

**Figure 14. SHAP value combinations**

Local explanations of individual flagged sessions further illustrate SHAP's utility. For example, an anomalous session flagged by IF involved a nurse accessing a patient's record at 02:15 a.m., 18 days after discharge, at a different provider than the primary provider, and following three accesses the previous day. SHAP assigned large positive contributions from provider mismatch, hour_of_day and days_since_discharge, resulting in a high anomaly score. In contrast, a false-positive session—where a physician accessed a record at 20:00 p.m. during an on-call shift at the primary provider but shortly after discharge—received a moderate anomaly score due to positive SHAP contributions from days_since_discharge and session_duration_sec but negative contributions from provider mismatch and hour_of_day. Such nuanced explanations help reviewers distinguish truly suspicious accesses from benign ones and could be incorporated into dashboards to support triage and training.

### 4.2. Discussion

#### 4.2.1. Summary of findings

The study sought to answer two research questions: (RQ1) What organisational and digital capabilities are required to adopt cross-provider AI-based anomaly detection, and (RQ2) How effective and explainable are lightweight anomaly-detection models in a simulated cross-provider setting? The scoping synthesis produced a four-pillar readiness checklist with ten concrete control items. Simulation results showed that simple rule sets provide high recall but at the cost of alert burden, while Isolation Forest reduces alerts but has low sensitivity. SHAP analysis revealed that provider mismatch and off-hours access are the dominant drivers of IF's anomaly scores, with interaction effects that amplify suspicion when multiple risk factors co-occur.

#### 4.2.2. Interpretation and comparison with prior work

The global and local SHAP explanations illuminate how unsupervised models implicitly learn heuristics similar to those advocated by domain experts. Provider mismatch emerged as the strongest indicator of anomalies, aligning with explanation-based auditing research which posits that most legitimate accesses can be explained by data (e.g., appointments); unexplained cross-provider accesses warrant scrutiny (Fabbri & LeFevre, 2011). The high SHAP values for off-hours and repeated access reflect patterns identified in previous security practice simulations—Yeng et al. (2021) reported that anomalous behaviours often involve accessing records outside typical shifts or at unusually high frequency. Thus, our findings corroborate earlier qualitative insights and provide quantitative evidence that such features are indeed discriminative.

However, the modest recall of IF in both datasets contrasts with reports from Tabassum et al. (2024), who found Isolation Forest to be among the top performers in smart-health contexts. The discrepancy may stem from the use of cross-provider contexts with rare anomalies and richer noise variables, highlighting the importance of contextual metadata and dataset characteristics. Deep-learning models like EHR-BERT may improve recall by capturing sequential patterns (Niu et al., 2024), but their interpretability and computational overhead pose challenges for routine auditing. Our simulation shows that a combination of simple rules (to guarantee coverage) and unsupervised models (to prioritise alerts) may provide a pragmatic balance, especially when coupled with interpretability tools.

#### 4.2.3. Implications for practice

The readiness checklist underscores that technical solutions must be supported by organisational infrastructure. Governance structures should establish policies for cross-provider audit logging, define what constitutes legitimate referrals and off-hours access, and assign responsibility for monitoring and escalation. Infrastructure investments should focus on standardising audit fields and enabling identity resolution across providers, as fragmentation undermines both detection and explanation (Ferreira et al., 2024). Workforce readiness requires training clinicians and compliance officers on the interpretation of algorithmic alerts and the significance of contextual features such as provider mismatch and off-hours. AI integration should include not only model deployment but also continuous monitoring and recalibration, with SHAP explanations integrated into dashboards to aid investigation and build trust. The strong influence of provider mismatch suggests that organisations should ensure referral information is captured in the audit log to reduce false positives.

#### 4.2.4. Limitations

Several limitations temper the generalisability of these findings. First, the simulation used synthetic audit logs; although synthetic data are necessary when real logs cannot be shared, they may not fully capture the heterogeneity of workflow patterns (Yeng et al., 2021). Second, only Isolation Forest and a simple rule set were evaluated. Other unsupervised algorithms (e.g., Local Outlier Factor, one-class SVM), semi-supervised approaches or sequence models may offer different trade-offs. Third, the feature set was limited; adding variables such as clinician specialty, access channel (web/mobile), or task context could enhance detection. Fourth, SHAP explanations are sensitive to feature scaling and model structure; while they provide valuable insights, they should be interpreted cautiously.

#### 4.2.5. Future research

Future work should validate these results on real multi-site audit logs and extend the feature set to include richer contextual signals. Experimenting with additional algorithms—especially deep-learning models and ensemble approaches—could explore whether higher recall can be achieved without sacrificing interpretability. Researchers should also investigate how SHAP explanations can be operationalised in practice: for example, by designing user interfaces that display local explanations and allow auditors to adjust thresholds dynamically. Finally, longitudinal studies that assess the impact of anomaly-detection systems on reviewer workload, clinician trust and organisational culture will be necessary to refine the readiness checklist and to ensure that technical advances translate into meaningful improvements in privacy and security.

## 5. Conclusion

Cross-provider exchange of electronic health records remains fragmented, creating blind spots for inappropriate access and insider misuse (Upadhyay & Hu, 2022). The present study examines the organisational capabilities required to adopt AI-based anomaly detection across shared data environments (RQ1) and evaluates the effectiveness and explainability of interpretable models in a simulated cross-provider setting (RQ2). Using a semi-systematic scoping synthesis combined with a simulation of synthetic audit logs, the work produces a ten-item readiness checklist spanning governance, infrastructure/interoperability, workforce and AI integration. In parallel, transparent rules are benchmarked against Isolation Forest (IF), with SHAP explanations used to characterise the feature interactions driving anomaly scores.

The findings advance current knowledge in three ways. First, the readiness checklist converts abstract notions of "digital readiness" into measurable controls and evidence artefacts, addressing persistent gaps in AI implementation guidance for health services (Nair et al., 2025). Second, the simulation shows that unsupervised models can reduce alert burden but may miss a material share of anomalies, whereas simple rules guarantee coverage at the cost of more false positives; a staged, hybrid approach therefore appears most practicable. Third, SHAP analysis provides case-level and feature-interaction explanations—highlighting cross-provider mismatch, off-hours access and repeated logins as dominant drivers—thus supporting transparent audit and triage (Fabbri & LeFevre, 2011).

Several limitations are acknowledged. Synthetic logs cannot fully reproduce real clinical workflows (cf. Yeng et al., 2021); only two algorithmic families are tested; and the contextual feature set is constrained.

Although deep-learning approaches may improve recall, their computational demands and reduced interpretability limit immediate applicability in operational settings (Nair et al., 2025). Future work should validate these results on multi-site audit logs, enrich contextual fields and algorithm breadth, measure reviewer workload and fairness, and link audit-log analytics to broader organisational-readiness processes.

Overall, effective anomaly detection in cross-provider EHR exchange is shown to depend not only on model accuracy but also on organisational readiness, transparent explanations and continuous governance. The proposed four-pillar framework and readiness checklist, combined with a rule-first then IF-augmented rollout, offer a practicable pathway toward trustworthy surveillance that protects patient privacy while enabling seamless data sharing.

## 6. Acknowledgement

This research benefited from the guidance and encouragement of Dr. Nagarajan Venkatachalam, whose timely feedback and supervision shaped the study's scope, methods and presentation.

Technical thanks are due to the open-source community whose tools enabled the simulation and analysis: Python/Jupyter, pandas, NumPy, scikit-learn, matplotlib, and SHAP. Generative-AI assistants (ChatGPT and GitHub Copilot) were used to accelerate coding of the Jupyter simulation environment and to suggest refactoring patterns; all code and text were reviewed, verified, and revised by the author, who bears full responsibility for the final content.